\documentclass[sigconf, review=false, screen=true, nonacm=true]{acmart}
\usepackage{wrapfig} 
\usepackage{xcolor} 
\usepackage{enumitem} 

\newcommand{\designreq}[1]{
    \textcolor{blue}{\textbf{DR#1}}
}

\AtBeginDocument{%
  }

\setcopyright{acmlicensed}
\copyrightyear{2025}
\acmYear{2025}
\acmDOI{} 
\acmConference[CHI '25]{the CHI Conference on Human Factors in Computing Systems}{April 28--May 01, 2025}{Yokohama, Japan}
\acmISBN{} 

\acmSubmissionID{8375}

\begin{document}

\title{Coolight: Enhancing Nighttime Safety for Urban Student Commuters}

\author{Mitsuka Kiyohara}
\authornote{Both authors contributed equally to this research.}
\email{mitsuka.kiyohara@mail.utoronto.ca}
\orcid{0009-0008-6420-9280} 
\author{Ethan Mondri}
\authornotemark[1]
\email{ethan.mondri@mail.utoronto.ca}
\orcid{0009-0008-8365-2560}
\affiliation{
  \institution{University of Toronto}
  \city{Toronto}
  \state{Ontario}
  \country{Canada}
}

\begin{abstract}
    Safety while walking alone at night is a key indicator of a citizen's well-being and a society's inclusiveness. 
    However, this is not equally felt across all demographic groups, especially for university students living in urban areas.
    We present Coolight, a mobile application designed to reduce stress and anxiety for nighttime walking through an interactive live map, real-time community incident reports, location sharing, and a route planner optimized for user safety. 
    Coolight's design was informed through interviews, questionnaires, and usability tests with university students and their friends and families in Toronto, Canada. 
    This paper describes the concept, research, design approach, and evaluation results of a solution addressing safety concerns urban commuters face at night.
\end{abstract}




\maketitle

\section{Introduction}

The feeling of safety while walking alone at night is a critical component of individual well-being and creating accessible and inclusive urban environments \cite{rausis_un, oecd, un_2015}.
Despite this being a common human activity, it continues to be a stressful and anxiety-inducing problem for people around the world. 
A 2024 WIN World Survey \cite{colwell_win_2024} found that 46\% of women and 26\% of men globally felt unsafe walking alone at night in their neighborhoods.
This problem is particularly felt among university students: survey research \cite{adtclergy2021} among 1,002 college students in Florida revealed that over 82\% of respondents expressed general concerns about personal safety, with 74\%  identifying walking home in the dark as an activity that makes them feel unsafe.
Urban areas exacerbate these concerns, as public spaces often become hotspots for crimes such as sexual violence or robberies to occur, significantly hindering urban mobility \cite{gc_2022}.
These safety challenges highlight an urgent need for solutions that address both physical security and the psychological reassurance of individuals navigating urban environments.
To address this problem, we present Coolight, a mobile application developed through a human-centered design process to enhance nighttime safety for students, families, and friends living in urban areas.
Coolight integrates safety-focused features such as optimized route planning, real-time community safety reports, and location sharing to empower users with proactive tools for safer commutes.
The main goal of Coolight is to contribute to the United Nations' Sustainable Development Goal (SDG) 11 of building inclusive, safe, resilient, and sustainable cities.

\section{Related Work}

There are a multitude of factors that affect perceived safety while walking at night. 
Prior research identified quiet streets, busy public spaces, and open empty spaces as contributing factors to feelings of unsafety \cite{uk_2022, lamour_2019}. 
García-Carpintero et al. \cite{gc_2022} also found lack of lighting and spaces with low influx of people are most perceived as dangerous by women living in cities.
Such research demonstrated the importance of designing solutions that prioritize environmental awareness and safety concerns posed by commuters, forming the foundation of our work.

Although diverse solutions exist to improve safety while walking, they address the issue in limited ways.
Navigation apps like Google Maps or Waze provide helpful features such as map navigation or street view to avoid getting lost in unfamiliar areas \cite{zhang_bandara_2024}; however, both prioritize efficiency over safety and are mostly (or fully, in Waze's case) driver-centered. 
Zhang and Bandara \cite{zhang_bandara_2024} also highlight that technology failure (e.g., incorrect routes,  risky routes) suggested by navigation apps makes pedestrians feel vulnerable or puts them in danger.
However, Waze is considered to produce higher user trust than Google Maps due to the higher degree of flexibility in information sharing \cite{navigation_trust}, which helped us recognize the importance of community-based mechanisms.
There also exist physical safety devices such as personal alarms or jewelry; however, these devices fall into gaps of being either too reactive or impractical for use \cite{jara-reinhold_analysis_2024}.
Another option for commuters is avoiding walking altogether by utilizing ride-hailing services or campus escort programs \cite{bedera_2015}; however, they are often limited by cost, availability, or scope, especially for off-campus commutes.
Therefore, existing solutions fall short of offering accessible and proactive tools.

\section{Formative Investigation}

To define and explore the scope of the problem, we conducted a series of formative studies with relevant stakeholder groups.
Our primary stakeholders were university students living in urban areas, while secondary stakeholders included their friends and family members.
These studies employed a combination of natural observations, questionnaires, and interviews, with questionnaires and interviews serving as the primary methods.

\subsection{Questionnaires}
We administered two online questionnaires for our primary and secondary stakeholders, with 30 and 20 respondents, respectively. 
The primary questionnaire was distributed to university students in the greater Toronto area, while the secondary questionnaire was disseminated by prompting participants from the primary questionnaire to share the survey with any of their friends or relatives.
Both questionnaires included single-choice, multiple-choice, and rating-scale questions to capture quantitative insights into stakeholder experiences.
The primary questionnaire focused on identifying positive and negative factors influencing perceived safety, current solutions utilized, and pain points faced during nighttime commutes.
The secondary questionnaire emphasized the concerns of friends and family, seeking to understand their experiences and challenges in supporting primary stakeholders.
Respondents to the primary questionnaire were primarily students in their early to late twenties, with 70\% aged between 20 and 22.
On the other hand, the secondary group comprised individuals from ages 19 to 62, with a mode of 6 people ages 20 and 22. 
The respondents were primarily men (60\%), and most lived in the same city as the commuters (55\%).

\subsection{Interviews}
To gain deeper qualitative insights, we conducted five anonymous, semi-structured interviews with four primary and one secondary stakeholder.
These virtual interviews aimed to uncover the lived experiences and emotional states of stakeholders while walking alone at night.
Key discussion points included environmental factors causing concern, reactive or adaptive strategies to perceived threats, and emotional responses during commutes.
Twelve prepared questions, informed by prior observations, guided the interviews, with additional follow-up questions used as needed to fully capture the interviewee's perspectives. 
To explore different perspectives, both male and female-identifying participants were recruited, with most being students, one recent graduate, and one significant other to a past student commuter.
All completed interviews were then transcribed, coded via affinity clustering, and later converted to actionable insights.

\begin{figure}
    \centering
    \begin{minipage}[b]{0.48\textwidth}
        \includegraphics[width=\textwidth]{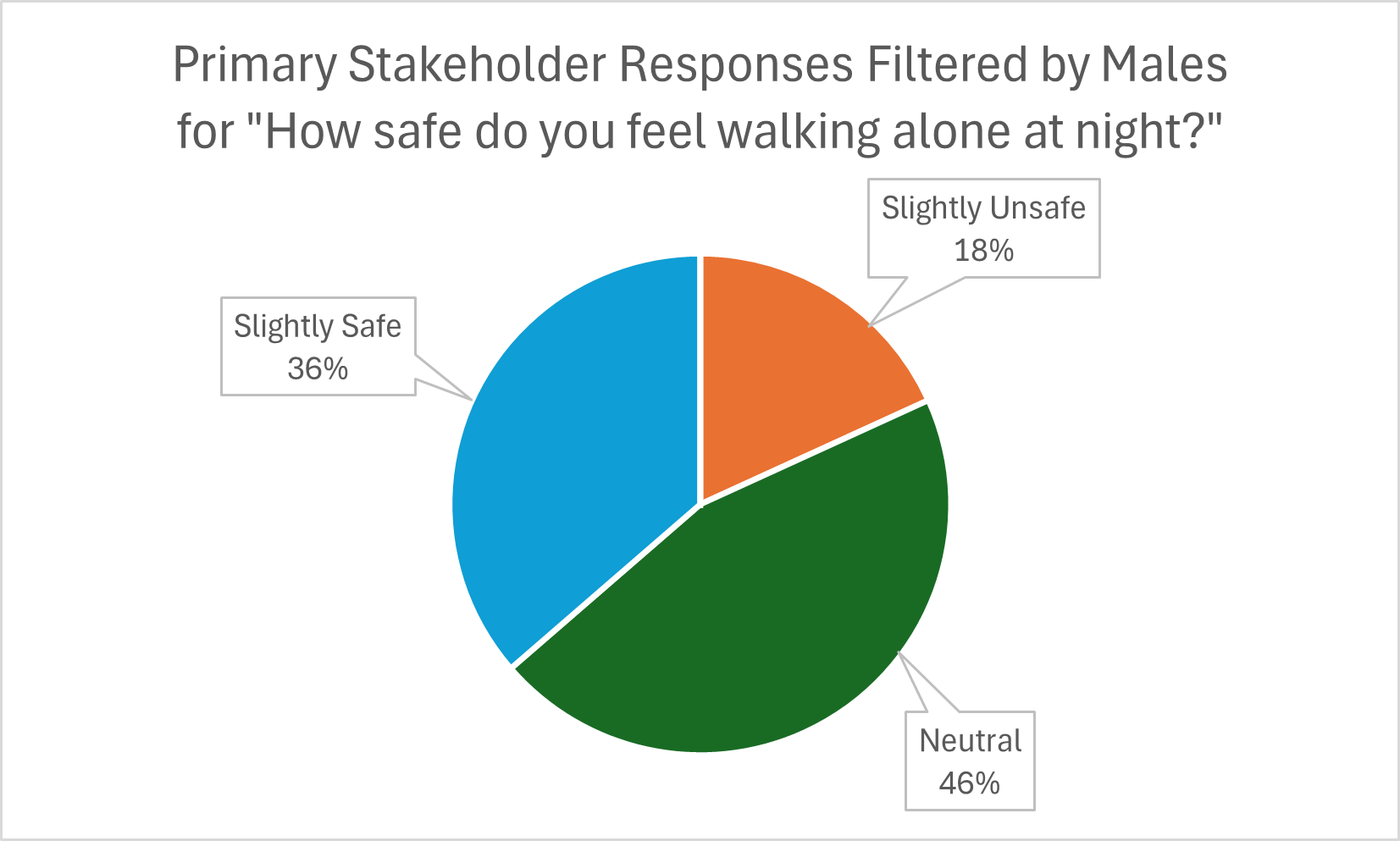}
        \caption{Primary Stakeholder Responses Filtered by Males for "How safe do you feel walking alone at night?"}
        \Description[Pie Chart of "how safe do you feel walking alone at night?" filtered by male respondents.]{Pie Chart of "how safe do you feel walking alone at night?" filtered by male respondents. The percentages are: 18\% "slightly unsafe", 46\% "neutral", and 36\% "slightly safe"}
        \label{fig:how_safe_male}
    \end{minipage}
    \hfill
    \begin{minipage}[b]{0.48\textwidth}
        \includegraphics[width=\textwidth]{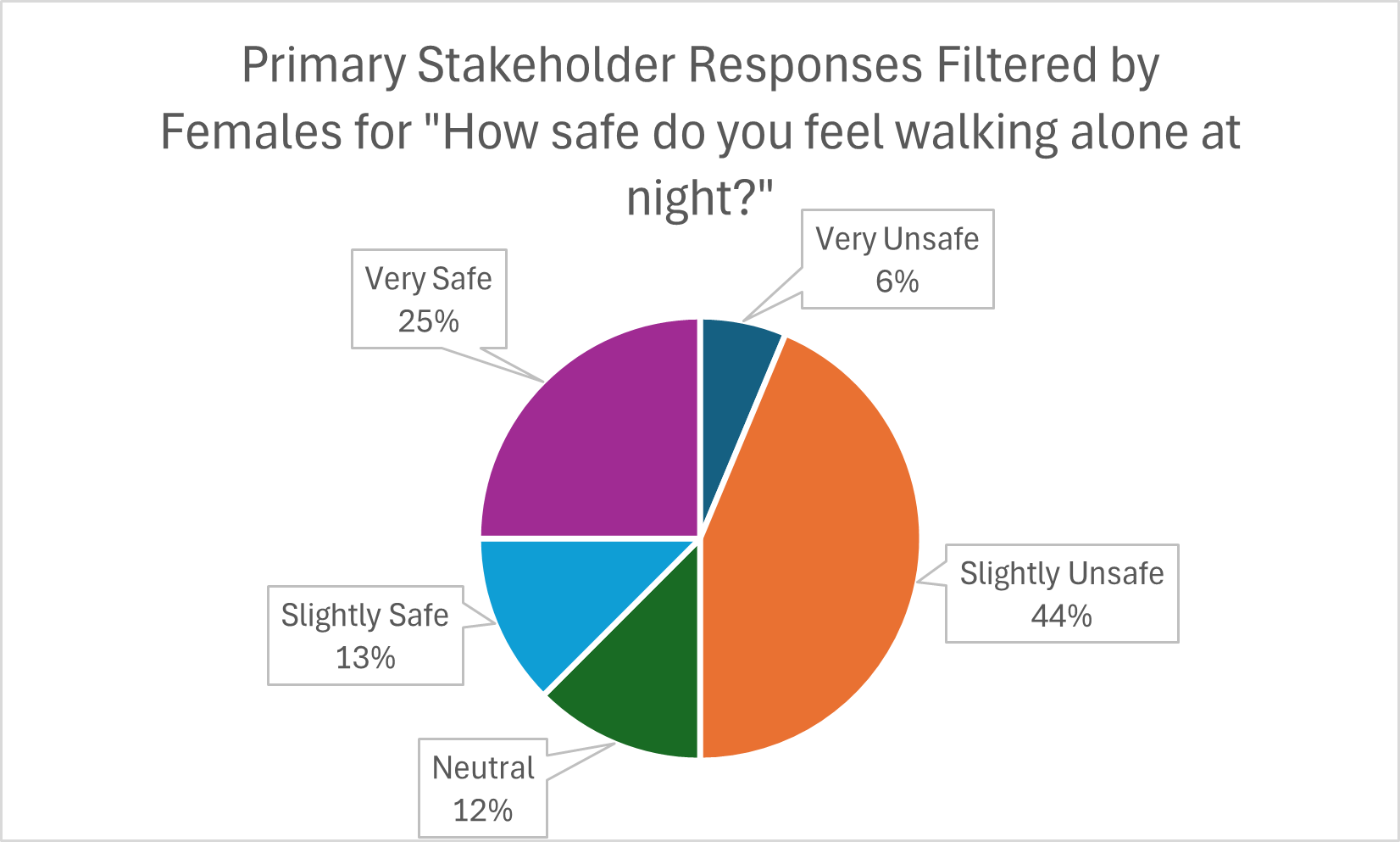}
        \caption{Primary Stakeholder Responses Filtered by Females for "How safe do you feel walking alone at night?"}
        \Description[Pie Chart of "how safe do you feel walking alone at night?" filtered by female respondents.]{Pie Chart of "how safe do you feel walking alone at night?" filtered by female respondents. The percentages are: 6\% "very unsafe", 44\% "slightly unsafe", 12\% "neutral", 13\% "slightly safe", and 25\% "very safe"}
        \label{fig:how_safe_female}
    \end{minipage}
\end{figure}

\subsection{Key Findings} \label{sec:key_findings}
From our formative studies, we developed three personas, an experience map, and job stories that describe the type of users who expressed the desire and need for support for staying safe while walking outside at night, alongside their practices and challenges when doing so.
The key insights discovered by these artifacts are outlined below:

\begin{itemize}
    \item Corroborating with existing research, women were more likely to feel unsafe walking alone at night, take more precautions, and be hyper-aware of their surroundings (Fig. \ref{fig:how_safe_male}, \ref{fig:how_safe_female}). 
    \item Key factors contributing to feelings of unsafety were poorly lit streets, the presence of individuals behaving erratically or under the influence of drugs or alcohol, and unfamiliarity with the area.
    \item Primary users preferred preventative measures to avoid dangerous confrontations altogether, such as walking with companions or crossing the street to avoid suspicious activity, rather than relying on reactive tools like personal safety devices.
    \item Primary and secondary users emphasized the value of staying alert and connected with trusted individuals through location sharing or texting before and after a walk. However, secondary users were frustrated with slow response times from the commuter and technical difficulties such as location sharing not updating. 
    \item Primary users found existing solutions like navigation apps or ride-hailing services inadequate to address their needs, citing the lack of safety-related features and high costs associated with ride-hailing for short distances.
\end{itemize}

From the analysis above, a clear need for an inclusive, accessible solution that prioritizes user safety through proactive measures was established.

\section{Design Process}

\subsection{Functional Design Requirements}
Building upon the job stories developed during user analysis, we established five key functional design requirements that depict what the end design intends to accomplish: 
\begin{itemize}
    \item [\designreq{1.}]  The design should be usable in a variety of ways (e.g., in advance, actively, through voice commands).
    \item [\designreq{2.}] The design should inform the features of a street (e.g., how well-lit it is, pedestrian traffic).
    \item [\designreq{3.}] The design should be easy and quick to use and understand.
    \item [\designreq{4.}] The design should integrate with existing mobile features and solutions.
    \footnote{Previously “The design should not prevent users from using their existing solutions.” Amended to be more specific about what the design should do, as well as to clarify the previous redundancy (no design should prevent users from using their existing solutions).}
    \item [\designreq{5.}] The design should provide user safety in a preventative manner rather than a responsive one.
\end{itemize}

\subsection{Ideation}

\subsubsection{Initial Concepts}
With these design requirements in mind,  we used the ``Crazy Eights'' method and referred to IDEO's 7 rules of brainstorming \cite{ideo_2020} to generate ideas.
After each round of brainstorming, we refined and combined recurring ideas into three initial design concepts: (i) a layered map, with each layer corresponding to a street feature that made people feel unsafe; (ii) a user report system (e.g., discussion board, blog) to inform and share safety concerns witnessed or experienced for others to view, comment, and vote; and (iii) a route planner prioritizing navigation for users' safety over speed. 
These three concepts were refined and combined again to form one solution using storyboards.
We created two high-level and three screen-level storyboards to visually understand how the product would interact with our target users (depicted through our formulated personas) in typical and edge-case scenarios.

\subsubsection{Final Concept}
As a result, a personal safety mobile application was conceived.
The primary interface (i.e., home screen) is a map with togglable/switchable layers, akin to the first concept.
However, the map also incorporates the second concept by showing report icons submitted by other users.
Users can access a route planning screen from the home screen, as outlined in the third concept.
Although not included in the three initial design concepts, we implemented a ``friends'' feature allowing users to optionally add contacts of friends and family to share their route and live location before leaving, as secondary stakeholders particularly emphasized the importance of maintaining contact during our formative studies (see Section \ref{sec:key_findings}).

\subsection{Paper (Low-Fidelity) Prototype}

\subsubsection{Paper Details}
Utilizing the devised final concept and screen-level storyboards, we created a low-fidelity (low-fi) prototype on paper to replicate user flows and allow usability evaluators to interact tangibly with the design.
This interactive prototype includes hand-drawn screens and ``interface widgets'' (e.g., buttons, checkboxes, keyboard, etc.) overlaid to simulate key interactions.

\subsubsection{Evaluations}
The low-fi prototype was then evaluated via heuristic evaluations and think-aloud evaluations.
Six experts assessed the prototype's usability based on Nielsen's ten heuristics \cite{nielsen_2024}.
Their feedback was later compiled and analyzed to identify each problem, its severity, and corresponding heuristic.
While the heuristic evaluations revealed the paper prototype to be intuitive, readable, and controllable, experts mainly highlighted concerns with the lack of consistent standards and help documentation, such as unclear iconography, lack of transparency in location sharing, and difficulties recognizing safety preferences in route optimization.
The prototype was updated to improve usability by redesigning the friends' screen for clearer location-sharing controls (see Fig. \ref{fig:location_sharing}) and more visible optimization preferences with floating tags and onboarding notes (see Fig. \ref{fig:route_planning}).

Subsequently, six university students participated in think-aloud evaluations to observe general user interactions and edge cases.
Each participant was provided with a real-world scenario and asked to perform three usability tasks with the paper prototype.
Each session was recorded across two camera angles to capture the participant's interactions, which were later transcribed, analyzed, and coded via affinity clustering.
While the paper prototype was generally found to be informative, feedback highlighted concerns about stigmatizing reporting categories and user privacy.
Overall, all participants expressed that such an app would be a reassuring tool when walking at night.

\subsection{Final (High-Fidelity) Prototype}
We introduce our final high-fidelity (hi-fi) prototype for a personal safety mobile application called Coolight, designed with color and interactions using Figma.

\subsubsection{Prototype Details}
\begin{figure}
    \centering
    \begin{minipage}[b]{0.19\textwidth}
        \includegraphics[width=\textwidth]{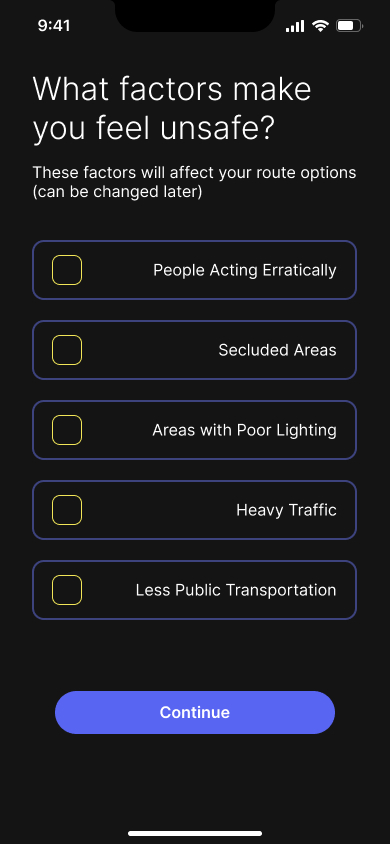}
        \caption{Onboarding}
        \Description[The onboarding screen of the described app, asking about factors that make the user feel unsafe.]{The onboarding screen of the described app, asking users to select one or more options of factors that make them feel unsafe including "People Acting Eratically", "Secluded Areas", "Areas with Poor Lighting", "Heavy Traffic", and "Less Public Transportation".}
        \label{fig:onboarding_factors}
    \end{minipage}
    \hfill
    \begin{minipage}[b]{0.19\textwidth}
        \includegraphics[width=\textwidth]{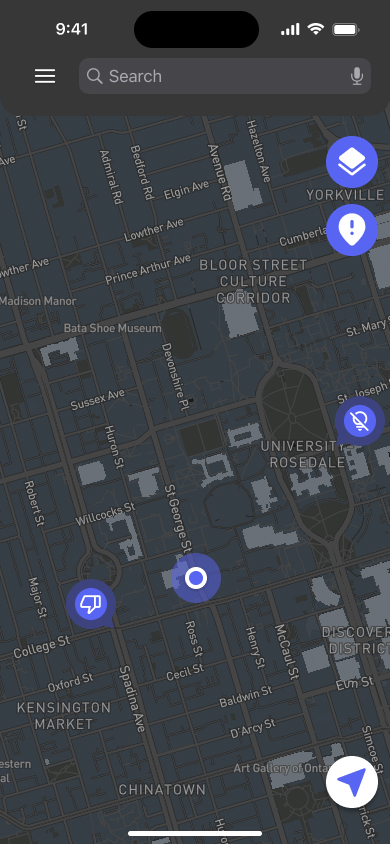}
        \caption{Live Map}
        \Description[A map interface.]{The live map screen of the described app, with a thumbs down icon visible at one intersection and a crossed out lightbulb at another.}
        \label{fig:live_map}
    \end{minipage}
    \hfill
     \begin{minipage}[b]{0.19\textwidth}
        \includegraphics[width=\textwidth]{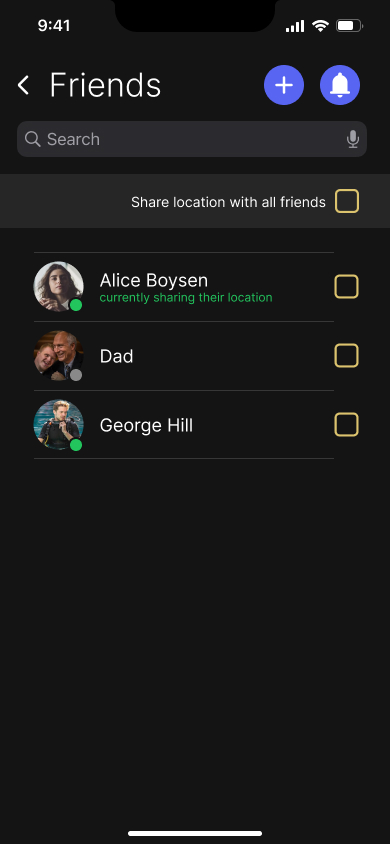}
        \caption{Location Sharing}
        \Description[A friends screen with three pictures and names, and tick boxes to share location with that friend]{A friends screen with three friends and tick boxes to share locations with them. One friend has a green label reading "currently sharing their location".}
        \label{fig:location_sharing}
    \end{minipage}
    \hfill
    \begin{minipage}[b]{0.19\textwidth}
        \includegraphics[width=\textwidth]{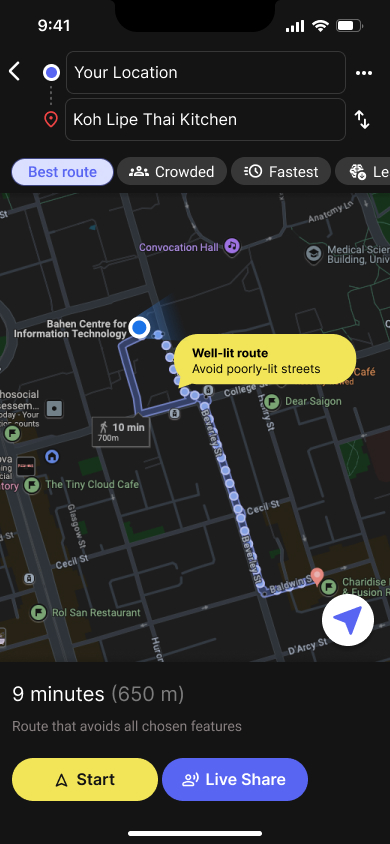}
        \caption{Route Planning}
        \Description[A screen showing the user a route from their location to a restaurant.]{A screen showing the user a route labeled "well-lit route, avoid poorly-lit streets" from their location to a restaurant.}
        \label{fig:route_planning}
    \end{minipage}
    \hfill
     \begin{minipage}[b]{0.19\textwidth}
        \includegraphics[width=\textwidth]{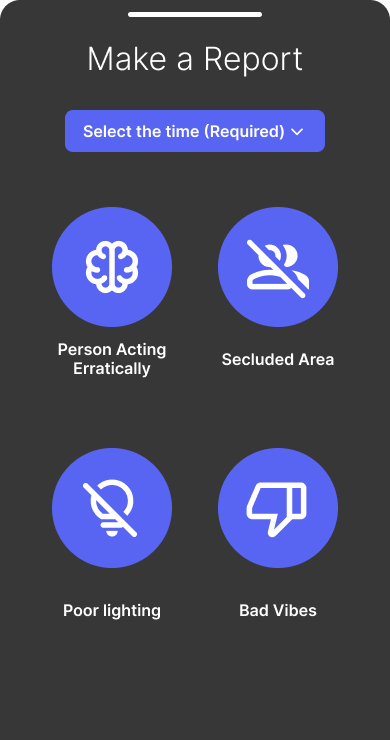}
        \caption{Reporting}
        \Description[A screen asking users to make a report and at what time.]{A screen with a title "Make a Report" and a time selection drop down, as well as four buttons for "Person Acting Erratically", "Secluded Area", "Poor Lighting", and "Bad Vibes".}
        \label{fig:incident_reporting}
    \end{minipage}
\end{figure}

\begin{itemize}
    \item \textit{The Live Map.}
    The live map is the central interface of the application and landing page after the onboarding process (Fig. \ref{fig:onboarding_factors}, \ref{fig:live_map}).
    Users can view real-time information about street conditions through switchable map layers (e.g., lighting levels) and user-submitted safety reports \textcolor{blue}{(\textbf{DR1}, \textbf{DR2)}}.
    By viewing reports ahead of time on the live map, users can avoid potentially unsafe areas before encountering them \textcolor{blue}{(\textbf{DR5})}.
    \item \textit{Incident Reporting.}
    Users can report safety concerns to share with the community both during and after their commute (Fig. \ref{fig:incident_reporting}).
    Reporting involves a simple, three-step process: tapping on a map location, setting a time, and choosing the report type \textcolor{blue}{(\textbf{DR3})}.
    The Quick Report button in the live map and navigation screen allows instant reporting by automatically selecting their current location, avoiding disruption during a user's walk \textcolor{blue}{(\textbf{DR3}, \textbf{DR1})}.
    \item \textit{Route-Planning.}
    Coolight offers navigation to users, prioritizing safety over speed based on user-selected preferences (Fig. \ref{fig:route_planning}).
    For advance planning, the route planner allows users to prepare for their commute beforehand, selecting paths that avoid certain characteristics they deem unsafe \textcolor{blue}{(\textbf{DR1})}.
    By providing a familiar navigation experience reminiscent of applications like Google Maps and Waze, Coolight fills a gap by focusing specifically on the walking commuter's safety \textcolor{blue}{(\textbf{DR4)}}.
    The route planning feature allows users to choose between routes tailored to their safety preferences, providing peace of mind \textcolor{blue}{(\textbf{DR5})}.
    \item \textit{Location Sharing.}
    When on a route, users can (optionally) share their location by selecting some of their previously added friends, to help reduce concerns of family and friends (Fig. \ref{fig:location_sharing}).
    Location sharing leverages existing mobile capabilities, allowing users to share their real-time location with friends and family for added safety \textcolor{blue}{(\textbf{DR4})}.
    Before leaving, users can share their planned route and real-time location with trusted contacts from the route planning page, ensuring that someone is aware of their journey and can respond promptly if needed \textcolor{blue}{(\textbf{DR5})}.
\end{itemize}

\subsubsection{Usability Testing}
A total of ten usability tests were conducted to evaluate the hi-fi prototype.
After the first six tests, we noticed all participants struggled with completing tasks due to technical issues with one of the hi-fi prototype's flows.
We immediately made changes to the prototype, and four additional tests were conducted on the updated version.
Conducting a second round of usability testing allowed for a stronger and smoother evaluation of the final hi-fi prototype.
The participants were all university students to best represent our target user group. 
We specifically selected individuals who had not interacted with the low-fi prototype for the data collection to be representative of new users.
For each test, participants were given the hi-fi prototype on a phone and asked to complete four usability tasks, designed to reflect realistic scenarios and evaluate how effectively participants could use Coolight to perform key actions on the hi-fi prototype.
The following tasks were adapted from the think-aloud evaluations (shortened for brevity):
\begin{itemize}
    \item [\textbf{Task 1.}] You are a student commuter with concerns about walking alone at night safely. After finishing a lecture, you would like to eat dinner at a restaurant. Show us how you would use this app to navigate to your destination and share your location with your dad Greg.
    \item [\textbf{Task 2.}] While walking, you notice a street has bad lighting. Show us how you report bad lighting at this location.
    \item [\textbf{Task 3.}] You’ve reached your destination, but you noticed someone acting erratically near the restaurant 1 minute ago. Please report this instance to help inform others using the app.
    \item [\textbf{Task 4.}] You're waiting for your friend John to show up at the restaurant. You know they are sharing their location with you on the app and want to check in on where they are. Demonstrate how you would use the app to view John’s route and location.
\end{itemize}
Participants’ screens and interactions were recorded, with note-takers documenting actions and feedback to ensure comprehensive data collection.
After all tasks were completed, participants were asked open-ended questions regarding the solution's usefulness, followed by an anonymous survey consisting of 10 System Usability Scale questions followed by three Likert questions.

To evaluate the usability of the hi-fi prototype, we used the ISO 9241-11 definition of usability and System Usability Scale (SUS).
To use ISO definition of usability, the key metrics recorded were \textit{effectiveness} (e.g., error count, task completion rate), \textit{efficiency} (e.g., time taken on tasks, perceived efficiency), and \textit{satisfaction} (e.g., satisfaction ratings, open-ended survey questions).

\subsubsection{Evaluation}
We summarize the key evaluation findings with respect to each usability metric.

\begin{itemize}
    \item \textit{Effectiveness.}
    On average, participants successfully completed 4.4 out of the 5 tasks, with less than two errors (e.g., missing taps, incorrect screen navigation) per task across both prototype versions.
    While Tasks 2 and 4 were the most successfully completed, Task 3 was the least successful largely due to two reasons: (i) participants forgot the need to select a time for their reports; and (ii) some participants exited from the live navigation page at the end of Task 2, which disrupted the tasks that followed.
    \item \textit{Efficiency.}
    Results showed that the final prototype was highly efficient.
    The longest tasks took participants just over half a minute, which is likely what led users to rate the prototype's efficiency as 4.9 out of 5.
    Although one participant mentioned that needing to select a time for reporting during Task 3 could be "somewhat annoying" for someone making many reports, this feature helps offset the accidental report with a default time (since reports can also be made retroactively).
    \item \textit{Satisfaction.}
    Post-survey results indicated the location-sharing and reporting features to be the most popular due to an added layer of comfort and access to useful information, respectively.
    While using Coolight, participants reported feeling calm and more safe when walking at night, with some preferring to use this solution over alternative navigation apps.
    On average, participants rated the hi-fi prototype at 4.4 out of 5 satisfaction.
    \item \textit{SUS.}
    After updating the hi-fi prototype, participants were overall more satisfied, with the System Usability Score improving from a B (86.25) to an A (91.25).
    This score change was also met with an increase in positive feedback from participants about their general experience interacting with the prototype.
    Overall, the System Usability Score across both versions indicated excellent usability (88.25).
\end{itemize}

\section{Discussion}
Below, we reflect upon the design and the lessons learned throughout the process.
While Coolight was primarily designed to address Sustainable Development Goal (SDG) 11 (especially Target 11.2),  we recognized that addressing safe mobility indirectly addressed multiple other SDGs.
For instance, by improving perceptions of safety while walking at night, Coolight directly contributes to SDG Target 16.1 (i.e., reduce all forms of violence and related death rates), which is measured by the population proportion that feels safe walking alone in their communities.
Coolight also indirectly supports SDG 3 by reducing stress and enhancing physical well-being through increased nighttime urban mobility.
Furthermore, since the problem of nighttime safety disproportionately affects women, our solution addresses SDG 5, promoting gender equality and ensuring that everyone feels safe in their environment.

In our efforts to create a positive and inclusive solution, we also reflected on ethical considerations to avoid unintended harm.
For example, we initially included a category for users to report the presence of homeless individuals, as this was identified as one of the top factors contributing to feelings of unsafety from our questionnaire responses.
However, after testing with users, we later recognized that this feature may contribute to stigmatization or othering of vulnerable populations.
This feedback prompted us to refine our design to focus on behaviors or environmental factors. 
Perceptions of safety are deeply personal and subjective.
While there is no definitive right or wrong answer, we realized the importance of approaching such topics with sensitivity and ensuring the design addresses user needs without reinforcing negative biases.

\section{Conclusion \& Next Steps}
This paper presents Coolight, a personal safety mobile application designed to enhance the nighttime safety of university students.
Through features such as an interactive, community-driven live map, real-time and retroactive incident reporting, safety-focused route planning, and location sharing, Coolight empowers users to make informed decisions and prioritize their well-being.
The design has been continuously refined with direct feedback from students and their family and friends at every stage of the design process to ensure a user-centered and ethically responsible solution.
Usability testing results demonstrate the app’s effectiveness, efficiency, and user satisfaction, validating its potential to address real-world safety concerns.
Beyond its immediate utility for students, Coolight lays the groundwork for further exploration of safety-driven technologies in urban mobility.
Our solution has the potential to serve as a valuable resource for city officials in data-driven analysis of walkability (e.g., street lighting conditions) while leveraging new technological integrations to promote safer and more accessible walking environments worldwide.
Nighttime safety is a universal challenge, and this project underscores the importance of leveraging thoughtful design and technology to address it.
By empowering users and fostering community collaboration, Coolight represents a step towards creating a safer, more inclusive future for urban commuters.

\begin{acks}
We would like to thank Anna Myllyniemi, Amrit Singh, and Zachary Lee for participating in the research and design process. We also share our deepest gratitude to Matthew Varona, Professor Khai N. Truong, and Professor Alex Mariakakis from the University of Toronto for their invaluable guidance. Lastly, we would like to thank all the participants throughout the development of this design.
\end{acks}

\bibliographystyle{ACM-Reference-Format}
\bibliography{CHI2025-bibliography}

\end{document}